\documentclass[a4paper,11pt]{article}
\usepackage{jheppub} 
\usepackage[T1]{fontenc} 
\RequirePackage[displaymath]{lineno} % Display line numbers
\usepackage{graphicx}% Include figure files
\usepackage{dcolumn}% Align table columns on decimal point
\usepackage{bm}% bold math
\usepackage{ltablex,booktabs}
\usepackage{overpic}
\usepackage{subfigure}
\usepackage{float}
\usepackage{color}
\usepackage{amsmath}
\usepackage{mathcomp}
\usepackage{mathrsfs}
\usepackage{multirow}
\usepackage{amssymb}
\usepackage{gensymb}
\usepackage{amsmath}
\usepackage{tabularx}
\usepackage{epstopdf}
\usepackage{ulem}
\usepackage{dutchcal}
\usepackage{indentfirst}

\title{Measurements of the absolute branching fractions of the doubly Cabibbo-suppressed decays $D^+\to K^+\pi^0$,  $D^+\to K^+\eta$ and  $D^+\to K^+\eta^{\prime}$}

\collaborationImg{\includegraphics[width=0.15\textwidth, angle=90]{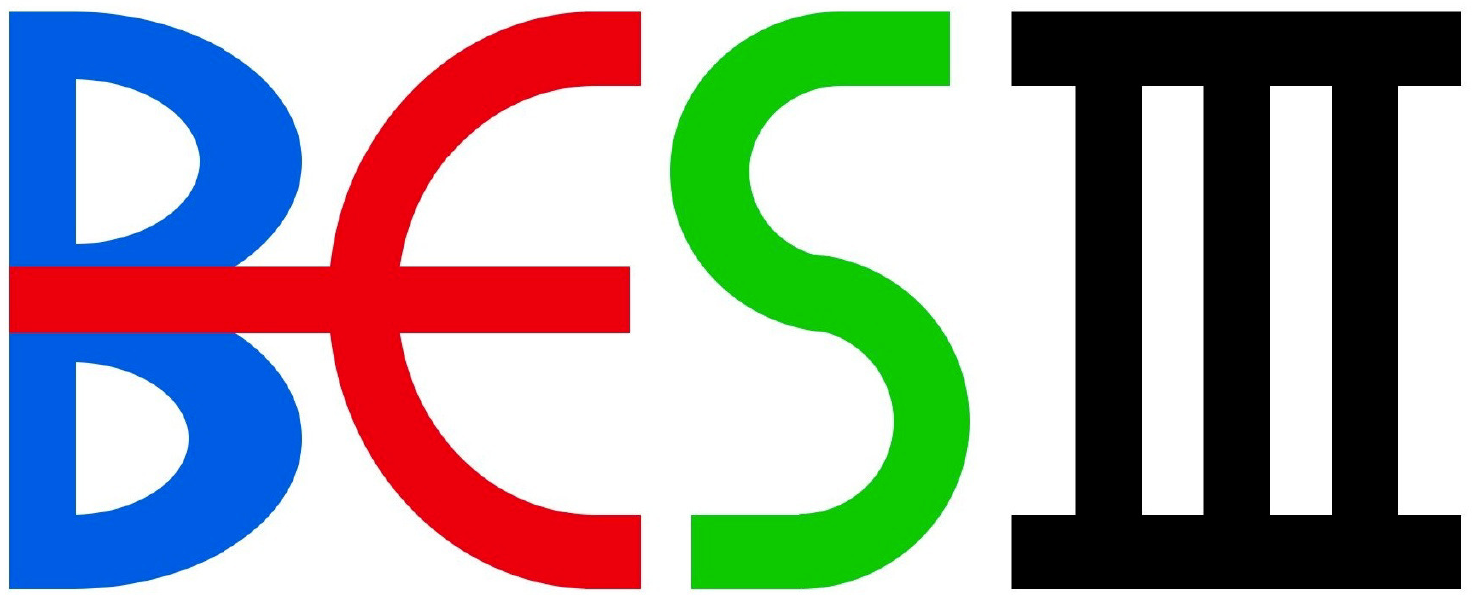}}

\collaboration{The BESIII Collaboration}

\emailAdd{besiii-publications@ihep.ac.cn}

\author{
M.~Ablikim$^{1}$, M.~N.~Achasov$^{4,c}$, P.~Adlarson$^{77}$, X.~C.~Ai$^{82}$, R.~Aliberti$^{36}$, A.~Amoroso$^{76A,76C}$, Q.~An$^{73,59,a}$, Y.~Bai$^{58}$, O.~Bakina$^{37}$, Y.~Ban$^{47,h}$
% ...
\\
\vspace{0.2cm}
(BESIII Collaboration)\\
\vspace{0.2cm} {\it
$^{1}$ Institute of High Energy Physics, Beijing 100049, People's Republic of China\\
% ...
$^{4}$ Budker Institute of Nuclear Physics SB RAS (BINP), Novosibirsk 630090, Russia\\
% ...
$^{36}$ Johannes Gutenberg University of Mainz, Johann-Joachim-Becher-Weg 45, D-55099 Mainz, Germany\\
$^{37}$ Joint Institute for Nuclear Research, 141980 Dubna, Moscow region, Russia\\
% ...
$^{47}$ Peking University, Beijing 100871, People's Republic of China\\
% ...
$^{58}$ Southeast University, Nanjing 211100, People's Republic of China\\
$^{59}$ State Key Laboratory of Particle Detection and Electronics, Beijing 100049, Hefei 230026, People's Republic of China\\
% ...
$^{73}$ University of Science and Technology of China, Hefei 230026, People's Republic of China\\
% ...
$^{76}$ University of Turin and INFN, (A)University of Turin, I-10125, Turin, Italy; (B)University of Eastern Piedmont, I-15121, Alessandria, Italy; (C)INFN, I-10125, Turin, Italy\\
$^{77}$ Uppsala University, Box 516, SE-75120 Uppsala, Sweden\\
% ...
$^{82}$ Zhengzhou University, Zhengzhou 450001, People's Republic of China\\

\vspace{0.2cm}
$^{a}$ Deceased\\
% ...
$^{c}$ Also at the Novosibirsk State University, Novosibirsk, 630090, Russia\\
% ...
$^{h}$ Also at State Key Laboratory of Nuclear Physics and Technology, Peking University, Beijing 100871, People's Republic of China\\
% ...
}

}

\abstract{Using $20.3\,\rm fb^{-1}$ of $e^+e^-$ collision data collected
	at a center-of-mass energy of 3.773\,GeV with the BESIII detector, we present  the measurements of the absolute branching fractions of the doubly Cabibbo-suppressed decays
	$D^+\to K^+\pi^0$, $D^+\to K^+\eta$ and $ D^+ \to K^+ \eta^{\prime}$ with the double-tag method, with significantly improved precision compared to the previous measurements. The statistical significance of each signal decay exceeds $10\sigma$.
	The branching fractions are determined to be ${\mathcal B}(D^+\to K^+ \pi^0) = (1.45 \pm 0.06 \pm 0.08)\times 10^{-4}$, ${\mathcal B}(D^+\to K^+ \eta) = (1.17 \pm 0.10 \pm 0.03)\times 10^{-4}$ and ${\mathcal B}(D^+\to K^+ \eta^{\prime}) = (1.88 \pm 0.15 \pm 0.11)\times 10^{-4}$, where the first uncertainties are statistical and the second systematic.
	The branching fractions of $D^+\to K^+\eta$ and $ D^+ \to K^+ \eta^{\prime}$ are consistent with the world average values. The reported branching fraction of $D^+\to K^+\pi^0$ deviates with the world average value by 3$\sigma$.}

    
\begin{document}
%\linenumbers
\maketitle
\flushbottom

\newpage
\section{Introduction}
Studies of doubly Cabibbo-suppressed (DCS) decays of charmed mesons are important for the understanding of charmed hadron dynamics. In theory, the branching fractions~(BFs) of the two-body decays $D \to PP$ and $D \to VP$, where $V$ and $P$ denote vector and pseudoscalar mesons, respectively, can be calculated by incorporating quark SU(3)-flavor symmetry and symmetry breaking as well as charge-parity ($CP$) violation~\cite{ref:the1,ref:the4,ref:the6,ref:the10,ref:the11,ref:the12}. These decay amplitudes are typically decomposed using topology amplitudes~\cite{ref:the13}. Compared to Cabibbo-favored and singly Cabibbo-suppressed $D$ decays, experimental information on DCS decays $D\to PP$ remains limited. Improved measurements of the BFs of these decays are able to provide valuable insights into charmed meson decays \cite{ref:the1,ref:the4,ref:the6,ref:the10,ref:the11,ref:the12,ref:the13}.

Previous measurements by CLEO, Belle, BaBar, and BESIII~\cite{ref:cleo,ref:belle,ref:babar,ref:bes3all,ref:pdg2022} have determined the BFs of the DCS decays $D^+\to K^+\pi^0$, $ D^+ \to K^+ \eta$ and $D^+\to K^+\eta^\prime$. Both CLEO and BESIII measurements are based on
$\psi(3770)$ data with single-tag~(ST) method which suffers from high backgrounds,
while Belle and BaBar employed relative methods with data on and near the $\Upsilon(4S)$ resonance~\cite{ref:belle,ref:babar}.

This paper reports the measurements of the absolute BFs of the DCS decays $D^+\to K^+\pi^0$, $D^+\to K^+\eta$ and $D^+ \to K^+ \eta^{\prime}$ using the double-tag~(DT) method~\cite{mark3}, with significantly improved precision compared to the previous measurements. Charge-conjugated decays are implied. The obtained BFs of $D^+\to K^+\eta$ and $ D^+ \to K^+ \eta^{\prime}$ are consistent with the world average values, while the BF of $D^+\to K^+\pi^0$ deviates with the world average value by $3\sigma$. This work is performed by using 20.3~\,fb$^{-1}$ of $e^+e^-$ collision data~\cite{lum_bes3} collected with the BESIII detector
at a center-of-mass energy of $\sqrt s=$ 3.773~GeV.
This energy corresponds to the $\psi(3770)$ resonance, which predominantly decays
into $D\bar D$ ($D$ denotes $D^0$ or $D^+$) pairs.
The $D$ and $\bar D$ mesons are produced without accompanying hadron(s), thereby offering an ideal environment to study $D$ meson decays with the DT technique. 
%The $CP$ asymmetry of $D^\pm \to K^\pm\pi^\pm\pi^\mp\pi^0$ is also presented.

\section{BESIII detector and Monte Carlo simulation}
The BESIII detector~\cite{ref:detector} records symmetric $e^+e^-$ collisions provided by the BEPCII storage ring~\cite{ref::collider}
in the center-of-mass energy range from 1.84 to 4.95~GeV,
with a peak luminosity of $1.1 \times 10^{33}\;\text{cm}^{-2}\text{s}^{-1}$
achieved at $\sqrt{s} = 3.773\;\text{GeV}$. 
The cylindrical core of the BESIII detector covers 93\% of the full solid angle and consists of a helium-based
multilayer drift chamber~(MDC), a plastic scintillator time-of-flight
system~(TOF), and a CsI(Tl) electromagnetic calorimeter~(EMC),
which are all enclosed in a superconducting solenoidal magnet
providing a 1.0~T magnetic field. The solenoid is supported by an
octagonal flux-return yoke with resistive plate counter muon identification modules interleaved with steel. The charged-particle momentum resolution at $1~{\rm GeV}/c$ is $0.5\%$, and the ${\rm d}E/{\rm d}x$ resolution is $6\%$ for electrons from Bhabha scattering. The EMC measures photon energies with a resolution of $2.5\%$ ($5\%$) at $1$~GeV in the barrel (end cap)
region. The time resolution in the TOF barrel region is 68~ps, while that in the end cap region was 110~ps. The end-cap TOF system was upgraded in 2015 using multi-gap resistive plate chamber technology, providing a time resolution of 60~ps~\cite{Tof1}, which benefits about 85\% of the data used in this analysis. Details about the design and performance of the BESIII detector are given in Ref. \cite{ref:detector}.

Simulated samples produced with a GEANT4-based~\cite{Geant4} Monte Carlo (MC) package, which
includes the geometric description of the BESIII detector and the
detector response, are used to determine the detection efficiency
and to estimate backgrounds. The simulation includes the beam
energy spread and initial state radiation (ISR) in the $e^+e^-$
annihilations modeled with the generator {\sc
	kkmc}~\cite{kkmc}.
The signal decays $D^+\to K^+\pi^0$, $D^+\to K^+\eta$ and $D^+ \to K^+ \eta^{\prime}$~are simulated to distribute uniformly in the available phase space. The decays $\eta^{\prime} \to \pi^+ \pi^- \eta$ and $\eta^{\prime} \to \pi^+ \pi^- \gamma$ are simulated by using a specific generator developed with the amplitude analysis result~\cite{evtgen}.
The background is studied using an inclusive MC sample that consists of the
production of $D\bar{D}$
pairs with consideration of quantum coherence for all neutral $D$
modes, the non-$D\bar{D}$ decays of the $\psi(3770)$, the ISR
production of the $J/\psi$ and $\psi(3686)$ states, and the
continuum processes incorporated in {\sc kkmc}.
The known decay modes are modeled with {\sc
	evtgen}~\cite{evtgen} using the known BFs taken from the
Particle Data Group (PDG)~\cite{ref:pdg2022}, while the remaining unknown decays
from the charmonium states are modeled with {\sc
	lundcharm}~\cite{lundcharm}. Final state radiation
from charged final state particles is incorporated with the {\sc
	photos} package~\cite{photos}.
\section{Measurement method and single tag yields }
Signal $D^+$ decays are reconstructed in events where $D^-$ decays are also reconstructed  through three hadronic decay modes: $D^-\to K^+\pi^-\pi^-$, $D^-\to K^0_S\pi^-$, and $D^-\to K^+\pi^-\pi^-\pi^0$. The $K^0_S$ and $\pi^0$ candidates are reconstructed via $K^0_S\to\pi^+\pi^-$ and $\pi^0\to\gamma\gamma$, respectively. Selection criteria for $K^\pm$, $\pi^\pm$, $K^0_S$ and $\pi^0$ follow Ref.~\cite{newref1}.  If a $D^-$ meson is found, it is referred to as an ST candidate. An event in which a signal $D^+$ decay and an ST $D^-$ are simultaneously reconstructed is referred to as a DT event.
The BF of the signal decay is determined as
\begin{equation}
	\label{eq:br}
	{\mathcal B(\rm sig)} = \frac{N_{\rm DT}}{\sum\limits_{i=1}^{3}{N^i_{\rm ST}(\epsilon_{\rm DT}^i/\epsilon_{\rm ST}^i)}},
\end{equation}
where ${ N_{\rm DT}}$ is the number of DT events, $\epsilon_{\rm DT}^i$ is the efficiency of detecting DT events including the sub-BFs, $N_{\rm ST}^i$ is the number of ST $D^-$ mesons, $\epsilon_{\rm ST}^i$ is the corresponding ST efficiency for the $i$-th tag mode.

\begin{figure*}[htp]
	\centering
	\includegraphics[width=1.0\linewidth]{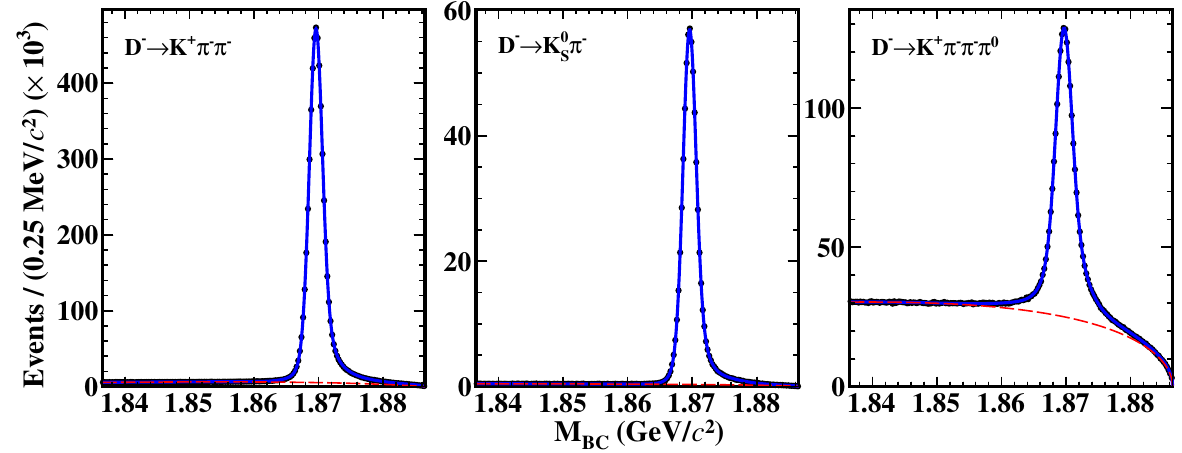}
	\caption{
		Fits to the $M_{\rm BC}$ distributions of the ST $D^-$ candidates. The dots with error bars are data, and the blue solid and red dashed curves are the fit results and the fitted backgrounds, respectively.}
	\label{fig:STfit}
\end{figure*}

The tagged $D^-$ mesons are selected using two variables: the energy difference
\begin{equation}\label{def_delE}
	\Delta E_{\rm tag} \equiv E_{D^-} - E_{\rm b},
\end{equation}
and the beam-constrained mass
\begin{equation}\label{def_mbc}
	M_{\rm BC}^{\rm tag} \equiv \sqrt{E^{2}_{\rm b}-|\vec{p}_{D^-}|^{2}},
\end{equation}
where $E_{\rm b}$ is the beam energy, and $\vec{p}_{D^-}$ and $E_{D^-}$ candidates are the momentum and the energy of the $D^-$ candidate in the $e^+e^-$ rest frame. For each tag mode, if there are multiple combinations, the one with the minimum $|\Delta E_{\rm tag}|$ is retained for further analysis.
For the tag modes $D^- \to K^+ \pi^- \pi^-$, $D^- \to K^0_S \pi^-$ \rm{~and} ~$D^- \to K^+ \pi^- \pi^- \pi^0$, the tagged $D^-$ candidates are required to satisfy
$\Delta E_{\rm tag}\in(-25,\, 24),~(-25,\, 25) \rm{~and} ~(-57,\, 46)$ \, MeV, respectively.
The yields of ST $D^-$ mesons are obtained from maximum likelihood fits to the $M_{\rm BC}^{\rm tag}$ distributions of the accepted ST candidates, as shown in Table \ref{tab:ST}. 
%Ref.~\cite{newref1}.
The fit results are shown in Fig.~\ref{fig:STfit}. The long tails of high side in each figure are mainly due to the ISR and final state radiation effects; while that in low side for Fig.~\ref{fig:STfit} (c) is mainly due to energy loss of photons. The total ST $D^-$ yield is ${N_{{\rm ST}}=}(8131.7\pm3.3)\times 10^3$.

\begin{table}
	\centering
	\caption{
		Requirement on $\Delta E$, ST $D^{-}$ yields in data and ST efficiencies ($\epsilon_{\rm ST}$)~\cite{newref1}.
	}
	\label{tab:ST}
	\centering
	\begin{tabular}{lccc}
		\hline	
		Tag mode &$\Delta E$ (MeV)&$N_{\rm{ST}}$ ($\times 10^3$)&$\epsilon_{\rm ST}(\%)$\\
		\hline
		$K^+\pi^-\pi^-$     &$[-25, 24]$&$5526.6\pm2.5$&$51.07\pm0.01$     \\
		$K^0_S\pi^-$&$[-25, 26]$&$~~656.5\pm0.8$&$51.42\pm0.01$ \\
		$K^+\pi^-\pi^-\pi^0$&$[-57, 46]$&$1740.2\pm1.8$&$24.53\pm0.01$  \\
		\hline		
	\end{tabular}
\end{table}

The signal $D^+$ candidates are reconstructed from the particles not used in the tagged $D^-$ reconstruction. The $K^0_S$, $\pi^0$ and $\eta$ candidates are reconstructed via the decays $K^0_S\to\pi^+\pi^-$ , $\pi^0\to\gamma\gamma$ and $\eta \to \gamma \gamma$, respectively. The selection criteria for $K^\pm$, $\pi^\pm$, $K^0_S$, $\pi^0$ and $\eta$ are the same as those in Ref.~\cite{newref1}. The $\eta^{\prime}$ signal regions are defined as $|M_{\pi^+\pi^-\eta}-0.958|<0.012$ GeV/$c^{2}$ and  $|M_{\pi^+\pi^-\gamma}-0.958|<0.020$ GeV/$c^{2}$. The corresponding sideband regions are $0.020<|M_{\pi^+\pi^-\eta}-0.958|<0.044$ GeV/$c^{2}$ and $0.030<|M_{\pi^+\pi^-\gamma}-0.958|<0.070$ GeV/$c^{2}$. 

The energy difference ($\Delta E_{\rm sig}$) and beam-constrained mass ($M_{\rm BC}^{\rm sig}$) are used to select signal $D^+$ candidates. These quantities are calculated similarly to Eqs.~(\ref{def_delE}) and~(\ref{def_mbc}), respectively, with $D^-$ replaced by $D^+$. If there are multiple combinations, primarily due to incorrect $\pi^0$ reconstruction, the combination with the minimum $|\Delta E_{\rm sig}|$ is retained for further analysis.
The signal side is required to satisfy $\Delta E_{\rm sig}\in (-72,49), (-43,41), (-32,31)$ and $(-43,40)~{\rm MeV}$~for $D^+\to K^+ \pi^0$, $D^+\to K^+ \eta$, $D^+\to K^+\eta^{\prime}_{\pi^+\pi^-\eta}$ and $D^+\to K^+\eta^{\prime}_{\pi^+\pi^-\gamma}$.

The $D^+\to K^+\pi^0$ candidates are required to have no additional $\pi^0$ meson to suppress the peaking background from $D^+ \to \pi^+ \pi^0 \pi^0$. To further reject possible $K^0_S$ background, a veto window $|M_{\pi^+\pi^-}-0.498|>0.012$~GeV/$c^2$ is applied for $D^+ \to K^+ \eta^{\prime}(\pi^+ \pi^- \gamma)$ candidates.

To suppress non-$D^+D^-$ events, such as $\psi(3770)\to D^0\bar{D}^0\to K^+ \pi^- K^0_S \eta$ for $D^+\to K^+\pi^0$ and $D^+\to K^+\eta$ channels, $\psi(3770)\to D^0\bar{D}^0\to K^+ \pi^- K^0_S \eta^{\prime}$ and $\psi(3770)\to D^0\bar{D}^0\to K^+ K^+ \pi^+ \pi^- \pi^- \pi^-$ for $D^+\to K^+\eta^{\prime}_{\pi^+\pi^-\eta}$ and $D^+\to K^+\eta^{\prime}_{\pi^+\pi^-\gamma}$ channels, respectively, The opening angle between the $D^+$ and $D^-$ candidates is required to be greater than $160^\circ$.
The left panels of Figs.~\ref{2Dfit} and \ref{2Dfit2} show the $M_{\rm BC}^{\rm tag}$ versus $M_{\rm BC}^{\rm sig}$ distributions of the accepted candidates for $D^+ \to K^+ \pi^0$, $D^+ \to K^+ \eta$, $D^+\to K^+\eta^{\prime}_{\pi^+\pi^-\eta}$ and $D^+\to K^+\eta^{\prime}_{\pi^+\pi^-\gamma}$ in data.

\section{Signal yields and Branching  fractions}
\begin{figure*}
	\centering
	\subfigure[]{\includegraphics[width=1.0\linewidth]{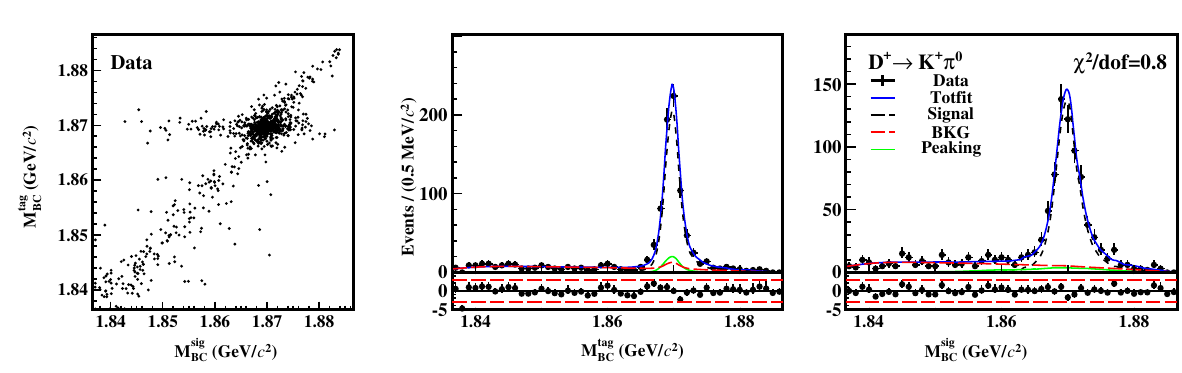}}
	\subfigure[]{\includegraphics[width=1.0\linewidth]{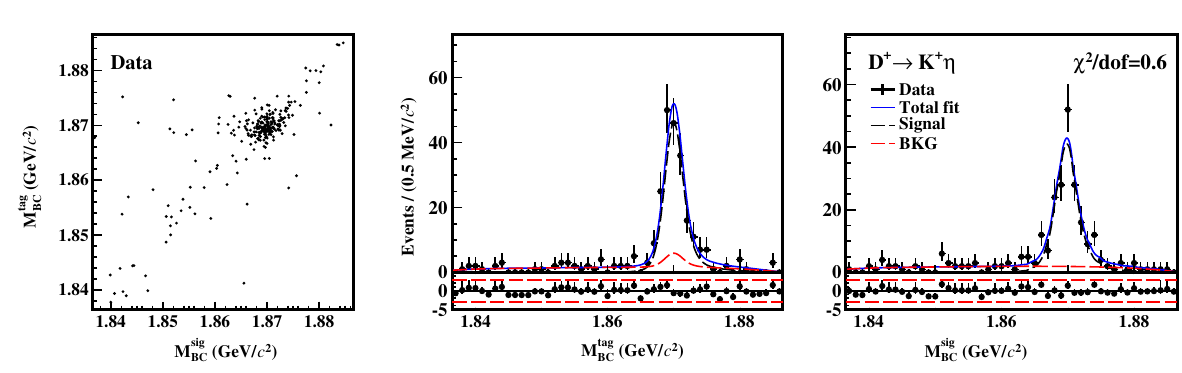}}
	\caption{
			The distributions of $M_{\rm BC}^{\rm tag}$ versus $M_{\rm BC}^{\rm sig}$
			and the projections on $M_{\rm BC}^{\rm sig}$ and
			$M_{\rm BC}^{\rm tag}$ of the 2D fits to the DT candidate events of (a) $D^+\to K^+\pi^0$ and (b) $D^+\to K^+ \eta$ in data sample with three tags. For both sub-figures (a) and (b), in the right two columns, the dots with error bars are data, the blue solid curves are the fit results, the black dashed lines are the signals, and the red dashed lines are the BKG I, II and III. For sub-figure (a), the green dashed line is fixed peaking background.}
	\label{2Dfit}
\end{figure*}

\begin{figure*}
	\centering
	\subfigure[]{\includegraphics[width=1.0\linewidth]{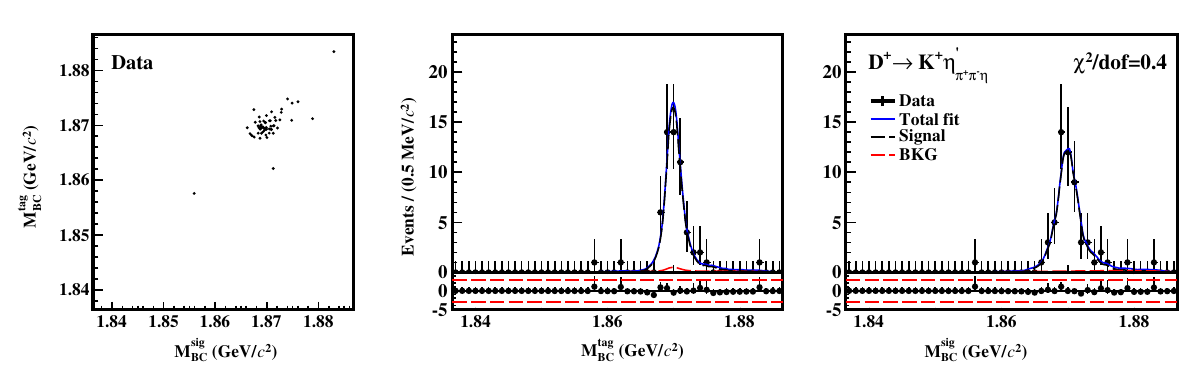}}
	\subfigure[]{\includegraphics[width=1.0\linewidth]{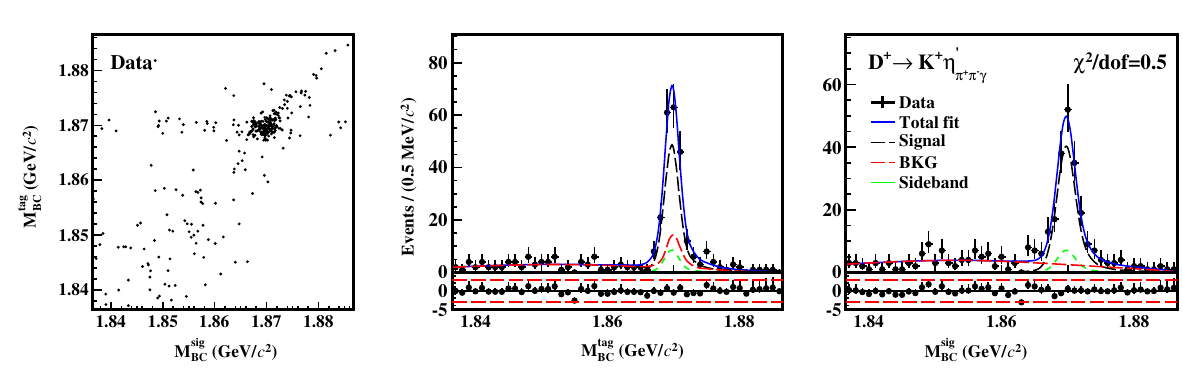}}
	\subfigure[]{\includegraphics[width=1.0\linewidth]{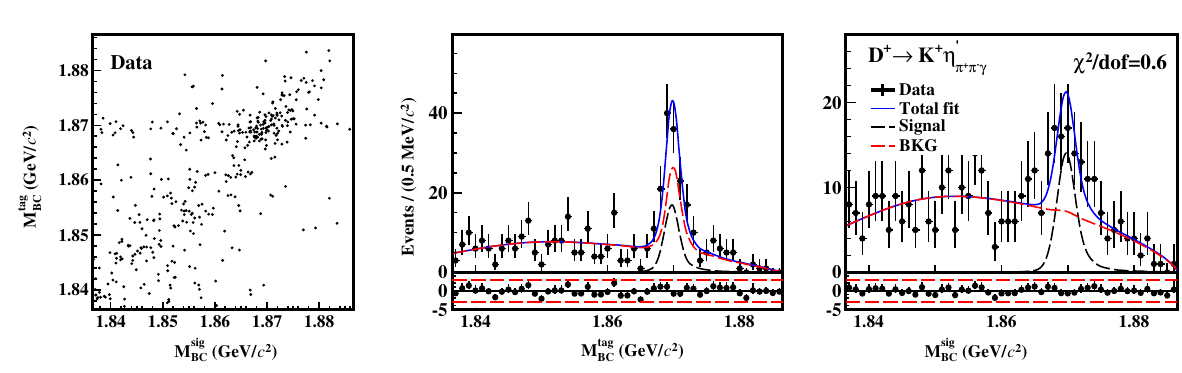}}
	\caption{
		The distributions of $M_{\rm BC}^{\rm tag}$ versus $M_{\rm BC}^{\rm sig}$ and the projections on $M_{\rm BC}^{\rm sig}$ and
		$M_{\rm BC}^{\rm tag}$ of the 2D fits to the DT candidate events of
		(a) $D^+\to K^+\eta^\prime(\pi^{+}\pi^{-}\eta)$,
		(b) $D^+\to K^+\eta^\prime(\pi^{+}\pi^{-}\gamma)$ in the $\eta^\prime$ signal region and
		(c) $D^+\to K^+\eta^\prime(\pi^{+}\pi^{-}\gamma)$ in the $\eta^\prime$ sideband region in data sample with three tags.
		In the fit of (c), the parameters of the smeared Gaussian function have been fixed to those in the $\eta^\prime$ signal region. For all three sub-figures (a), (b) and (c), in the right  two columns, the dots with error bars are data, the blue solid curves are the fit results, the black dashed lines are the signals, and the red dashed lines are the BKG I, II and III. For sub-figure (b), the green dashed line is the normalized sideband events. For sub-figure (a), due to the low backgrounds, the black curves overlap with blue curves. }
	\label{2Dfit2}
\end{figure*}

The yields of DT events are extracted from a two-dimensional (2D) extended unbinned maximum likelihood fit to the corresponding distribution of $M_{\rm{BC}}^{\rm{tag}}$ versus $M_{\rm{BC}}^{\rm{sig}}$ ~\cite{cleo-2Dfit,bes2Dfit1,bes2Dfit2}. The signal shape is described by a 2D probability density function (PDF) from the MC simulation after convolved with a Gaussian resolution function with parameters derived from data. For various background components, the individual PDFs are constructed as follows:~\cite{cleo-2Dfit,bes2Dfit1,bes2Dfit2}
\begin{itemize}
	\item BKGI: $b(x)\cdot c_y(y;E_{\rm b},\xi_{y}) + b(y)\cdot c_x(x;E_{\rm b},\xi_{x})$,
	\item BKGII: $c_z(z;\sqrt{2}E_{\rm b},\xi_{z}) \cdot g(k;0,\sigma_k)$,
	\item BKGIII: $c_x(x;E_{\rm b},\xi_{x}) \cdot c_y(y;E_{\rm b},\xi_{y})$.
\end{itemize}
Here, $x=M_{\rm BC}^{\rm tag}$, $y=M_{\rm BC}^{\rm sig}$, $z=(x+y)/\sqrt{2}$, and $k=(x-y)/\sqrt{2}$.
The functions $b(x)$ and $b(y)$ are the one-dimensional signal shapes taken from the MC simulation.
The function $c_f$ is an ARGUS function~\cite{ARGUS} defined as
\begin{equation}
	c_f\left(f; E_{\rm b}, \xi_f\right) = A_f f (1 - \frac {f^2}{E_{\rm b}^2})^{\frac{1}{2}} e^{\xi_f (1-\frac {f^2}{E_{\rm b}^2})},
\end{equation}
where $f$ denotes $x$, $y$, or $z$, $E_{\rm b}$ is fixed at 1.8865 GeV, $A_f$ is a normalization factor,
and $\xi_f$ is a fit parameter.
The function $g(k;\sigma_k)$ is a Gaussian distribution with a mean of zero and a standard deviation $\sigma_k=\sigma_0 \cdot(\sqrt{2}E_{\rm b}-z)^p$,
where $\sigma_0$ and $p$ are free parameters in the fit. All other parameters are left free to vary. The three types of background are illustrated as one curve in Figs.~\ref{2Dfit} and \ref{2Dfit2}. For $D^+ \to K^+ \pi^0$, the peaking background yield, including $\pi^+\pi^0$ and $K^0_S\pi^+$, is fixed at the value estimated from the inclusive MC samples. Furthermore, for $D^+\to K^+\eta^\prime(\pi^+\pi^-\eta)$ and $D^+\to K^+\eta^\prime(\pi^+\pi^-\gamma)$, a simultaneous fit is performed to extract the BF as a shared parameter. The contribution from the $D^+\to K^+\eta^\prime(\pi^+\pi^-\gamma)$ sideband regions is used to estimate the numbers of non $\eta^\prime(\pi^+\pi^-\gamma)$ background events and is  re-normalized to the signal region in the fit procedure. The contribution from the $D^+\to K^+\eta^\prime(\pi^+\pi^-\eta)$ sideband regions is neglected since only one event survives after event selection. 

The spectra in the middle and right columns in Figs.~\ref{2Dfit} and \ref{2Dfit2} show
the projections on $M_{\rm BC}^{\rm tag}$ and $M_{\rm BC}^{\rm sig}$ of the 2D fits to data.
For each signal decay, the statistical significance is evaluated as $\sqrt{-2\ln(\mathcal L_0/\mathcal L_{\rm max})} $,
where $\mathcal L_{\rm max}$ is the maximum likelihood of the nominal fit and $\mathcal L_0$ is the likelihood of the fit without including the signal. The change of the number of degrees of freedom is assumed to be 5 for $K^+ \pi^0$ and $K^+\eta$ decay mode, representing the signal yield. The change of the number of degrees of freedom is assumed to be 7 for $K^+ \eta^{\prime}$, representing  the signal yield of $K^+ \eta^{\prime}(\pi^+\pi^-\eta)$, the signal yield of $K^+ \eta^{\prime}(\pi^+\pi^-\gamma)$ and the shared parameter for the BF of $D^+ \to K^+\eta^{\prime}$.
The statistical significance of each signal decay exceeds 10$\sigma$.
The detailed ST, DT and DT/ST efficiencies are shown in Table~\ref{tab:doubletageff}.
The signal yields in data, the detection efficiencies, and the obtained BFs of each signal decay are summarized in Table~\ref{tab:result}. 
\begin{table}
	\centering
	\caption{
		The numbers used in the BF calculations. $N_{\rm DT}$ is the fitted number of the DT events in data, $\epsilon_{\rm DT}/\epsilon_{\rm ST}$ is the efficiency of detecting the signal decay in the presence of the ST $D^-$, which includes BFs of the intermediate states.~$\mathcal B_{\rm sig}$ is the BF of signal decay, where the uncertainties are statistical only.
	}
	\label{tab:result}
	\centering
	\begin{tabular}{cccc}
		\hline	
		Signal decay &$D^+\to K^+\pi^0$&$D^+\to K^{+}\eta$ &$D^+\to K^+\eta^{\prime}$\\
		\hline
		
		$N_{\rm DT}$                                &$629\pm28$  &$182\pm15$   &$214\pm17$     \\
		$\epsilon_{\rm DT}/\epsilon_{\rm ST}$ (\%)                      &$53.29\pm0.36$ &$19.20\pm0.13$  &$14.02\pm0.11$ \\
		$\mathcal B_{\rm sig}~(\times 10^{-4})$              &$1.45\pm0.06$  & $1.17\pm0.10$&$1.88\pm0.15$  \\
		\hline		
	\end{tabular}
\end{table}

\begin{table}[H]
	\centering
	\caption{The efficiencies  $\epsilon^{i}_{\rm{DT}}$ and $\epsilon^{i}_{\rm{DT}}/\epsilon^{i}_{\rm{ST}}$ which includes BFs of the intermediate states. The efficiencies are weighted by the corresponding ST yields in data. The uncertainties are statistical only.}
	\label{tab:doubletageff}
	
		\begin{tabular}{lcc}
			\hline
			Tag mode &    $\epsilon^{i}_{\rm DT}$~for $D^+\to K^+\pi^0~(\%)$ 
			 & $\epsilon^{i}_{\rm{DT}}/\epsilon^{i}_{\rm{ST}}$ for $D^{+}\to K^+\pi^0~(\%)$ \\ \hline
			$D^-\to K^+\pi^-\pi^-$                &$27.23\pm0.17$&$53.31\pm0.33$ \\
			$D^-\to K^{0}_{S}\pi^{-}$               &$27.75\pm0.17$&$53.97\pm0.33$ \\
			$D^-\to K^{+}\pi^{-}\pi^{-}\pi^{0}$                 &$12.84\pm0.12$&$52.36\pm0.49$ \\
			Weight &  &$53.29\pm0.36$ \\ \hline
			
				Tag mode 
			& $\epsilon^{i}_{\rm DT}$~for $D^+\to K^+\eta~(\%)$ & $\epsilon^{i}_{\rm{DT}}/\epsilon^{i}_{\rm{ST}}$ for $D^{+}\to K^+\eta~(\%)$ \\ \hline
			$D^-\to K^+\pi^-\pi^-$                   &$9.83\pm0.06$&$19.24\pm0.12$ \\
			$D^-\to K^{0}_{S}\pi^{-}$                 &$9.95\pm0.06$&$19.36\pm0.12$ \\
			$D^-\to K^{+}\pi^{-}\pi^{-}\pi^{0}$                 &$4.59\pm0.04$&$18.72\pm0.18$ \\
			Weight &  &$19.20\pm0.13$ \\ \hline
			
				Tag mode 
			& $\epsilon^{i}_{\rm DT}$~for $D^+\to K^+\eta^{\prime}~(\%)$ & $\epsilon^{i}_{\rm{DT}}/\epsilon^{i}_{\rm{ST}}$ for $D^{+}\to K^+\eta^{\prime}~(\%)$ \\ \hline
			$D^-\to K^+\pi^-\pi^-$                  &$7.28\pm0.05$&$14.26\pm0.10$ \\
			$D^-\to K^{0}_{S}\pi^{-}$                &$7.53\pm0.05$&$14.64\pm0.11$ \\
			$D^-\to K^{+}\pi^{-}\pi^{-}\pi^{0}$                 &$2.99\pm0.03$&$12.21\pm0.14$ \\
			Weight &  &$14.02\pm0.11$ \\ \hline
		\end{tabular}
	
\end{table}

\section{Systematic uncertainty}

With the DT method, most of the uncertainties related
to the ST selection  cancel out. The systematic uncertainties arise from the following
sources and are estimated relative to the measured BFs.

The uncertainty of the total ST $D^-$ yield
is due to the fit to the $M_{\rm BC}^{\rm tag}$ distributions
and is assigned as
0.3\%~\cite{newref1}. 

The tracking or particle identification~(PID) efficiencies of $K^+$ and $\pi^\pm$ are estimated by analyzing DT $D^+\to K^-\pi^+\pi^+$ versus $D^-\to K^+\pi^-\pi^-$ hadronic events. The systematic uncertainties in the tracking or PID are assigned as 1.0\% per track~\cite{ref:jianfengDCS}.

The systematic uncertainty of the $\gamma$ selection is estimated by studying the control sample of $J/\psi \to \pi^0 \pi^+ \pi^-$~\cite{ref:gammasel}.~The systematic uncertainty in the $\gamma$ selection is assigned as 1.0\% per photon.

The systematic uncertainty of the $\pi^0$ reconstruction has been studied by using the DT $\bar{D}^0\to K^+\pi^-$ versus $D^0\to K^-\pi^+\pi^0$ events. We assign the systematic uncertainty in the $\pi^0$ reconstruction to be 2.0\% per $\pi^{0}$~\cite{ref:jianfengDCS}. Due to limited $\eta$ sample, the systematic uncertainty of the $\eta$ reconstruction is assigned as 2.0\% per $\eta$ by referring to that of $\pi^0$.

The systematic uncertainty of the $\eta^{\prime}$ mass window selection has been studied in $D^+ \to \eta^{\prime} \ell^{+} v_{\ell}$~\cite{ref:etappan}, and computed
as 0.7.

The uncertainties of the quoted BFs of $\pi^0\to\gamma\gamma$, $\eta\to\gamma\gamma$, $\eta^{\prime}\to\pi^+ \pi^- \eta$ and $\eta^{\prime}\to\pi^+ \pi^- \gamma$ are 0.03\%, 0.5\%, 1.2\% and 1.4\%, respectively~\cite{ref:pdg2022}.

The systematic uncertainty of the $\Delta E^{\rm sig}$ requirement is estimated with the control samples of $D^+\to \pi^+\pi^0$, $D^+\to \pi^+\eta$, $D^+\to \pi^+\eta^{\prime}_{\pi^+\pi^-\eta}$ and $D^+\to \pi^+\eta^{\prime}_{\pi^+\pi^-\gamma}$. The differences in efficiencies between data and inclusive MC sample are taken as the systematic uncertainties.
They are 1.1\%, 0.4\%, 0.1\% and 0.1\% for $D^+\to K^+\pi^0$, $D^+\to K^+\eta$, $D^+\to K^+\eta^{\prime}_{\pi^+\pi^-\eta}$ and $D^+\to K^+\eta^{\prime}_{\pi^+\pi^-\gamma}$, respectively. 

The systematic uncertainties of the 2D fit are mainly due to signal shapes, background shapes, and peaking background. The uncertainties due to signal shapes are estimated by using Crystal-ball function. The uncertainties due to the background shapes are estimated by varying the endpoint by 0.2~MeV$/c^2$ for the ARGUS function. The systematic uncertainties due to peaking background are estimated by varying the BFs of the peaking background channels by $\pm 1\sigma$ and the data-MC difference of the peaking background channels including PID, tracking, $\pi^0$ reconstruction, $K^0_S$ reconstruction, $\Delta E^{\rm sig}$ cut and opening angle requirement~\cite{ref:pdg2022, lum_bes3, ref:jianfengDCS}. After adding these three uncertainties in quadrature, the systematic uncertainties of the 2D fit are assigned to be 3.9\%, 1.2\%, 4.8\% and 4.8\% for $D^+\to K^+\pi^0$, $D^+\to K^+\eta$, $D^+\to K^+\eta^{\prime}_{\pi^+\pi^-\eta}$ and $D^+\to K^+\eta^{\prime}_{\pi^+\pi^-\gamma}$, respectively. 

The systematic uncertainty of $K^0_S$ veto is estimated by examining the BFs after enlarging the veto window by $\pm 10$ and $\pm20$~MeV/$c^2$. The change of the re-measured BF, 1.8\%, is assigned for $D^+ \to K^{+} \eta^{\prime}$.

The systematic uncertainty of fit bias is estimated with 40 times fake data which include signal MC with input BFs and background derived from inclusive MC, each with an equivalent size to the data. The difference between the fit mean value of the 40 times fake data and the input value, 3.0\%, is taken as systematic uncertainty for $D^+ \to K^{+} \pi^{0}$. The systematic uncertainty is negligible for other signal decays.

The systematic uncertainties due to the $D^+ D^-$ opening angle requirement are estimated by using the control samples of $D^+\to \pi^+\pi^0$, $D^+\to \pi^+\eta$, $D^+\to \pi^+\eta^{\prime}_{\pi^+\pi^-\eta}$ and $D^+\to \pi^+\eta^{\prime}_{\pi^+\pi^-\gamma}$.~The differences in the acceptance efficiencies between data and MC simulation are assigned as individual systematic uncertainties, which are 1.3\%, 0.2\%, 2.8\% and 1.9\% for $D^+\to K^+\pi^0$, $D^+\to K^+\eta$, $D^+\to K^+\eta^{\prime}_{\pi^+\pi^-\eta}$ and $D^+\to K^+\eta^{\prime}_{\pi^+\pi^-\gamma}$, respectively.  

By adding each of the systematic uncertainties in quadrature, we obtain the total systematic uncertainties for $D^+\to K^+\pi^0$, $D^+\to K^+\eta$, $D^+\to K^+\eta^{\prime}_{\pi^+\pi^-\eta}$ and $D^+\to K^+\eta^{\prime}_{\pi^+\pi^-\gamma}$, respectively. All these systematic uncertainties are summarized in Table~\ref{tab:relsysuncertainties}.~~Since the $D^+\to K^+\eta^{\prime}_{\pi^+\pi^-\eta}$ and $D^+\to K^+\eta^{\prime}_{\pi^+\pi^-\gamma}$ are fitted simultaneously,  we consider the systematic uncertainties correlation of $D^+\to K^+\eta^{\prime}_{\pi^+\pi^-\eta}$ and $D^+\to K^+\eta^{\prime}_{\pi^+\pi^-\gamma}$. The correlated systematic uncertainties include the total ST $D^-$ yield, $K^{+}(\pi^\pm)$ PID or tracking, $\eta^{\prime}$ mass window and 2D fit; the total effect of these sources is 5.7\%. The un-correlated systematic uncertainties are from $\gamma$ selection, $\pi^{0} (\eta)$ reconstruction, quoted BFs, $\Delta E^{\rm sig}$, MC statistics, $K^0_S$ veto, fit bias and opening angle requirement; they are 2.7\% and 2.5\% for $D^+\to K^+\eta^{\prime}_{\pi^+\pi^-\eta}$ and $D^+\to K^+\eta^{\prime}_{\pi^+\pi^-\gamma}$, respectively.~The total systematic uncertainty is obtained to be 5.8\% by combining the systematic uncertainties of $D^+\to K^+\eta^{\prime}_{\pi^+\pi^-\eta}$ and $D^+\to K^+\eta^{\prime}_{\pi^+\pi^-\gamma}$ \cite{ref:comb}.

\begin{table*}[htp]
	\centering
	\caption{
		Relative systematic uncertainties (\%) in the BF measurements. The top and bottom parts are correlated and uncorrelated systematic uncertainties for $D^+\to K^+\eta^\prime_{\pi^+\pi^-\eta}$ and $D^+\to K^{+}\eta^\prime_{\pi^+\pi^-\gamma}$, respectively. The ... means that the uncertainty is negligible. }
	\label{tab:relsysuncertainties}
	\centering
	\resizebox{1.0\textwidth}{!}{
	\begin{tabular}{ccccc}
		\hline
		Uncertainty&$D^+\to K^+\pi^0$&$D^+\to K^+\eta$&$D^+\to K^+\eta^\prime_{\pi^+\pi^-\eta}$&$D^+\to K^{+}\eta^\prime_{\pi^+\pi^-\gamma}$\\
		\hline
		$N_{\rm tag}$                      &0.3&0.3&0.3&0.3\\
		$K^+(\pi^{\pm})$ tracking or PID                  &1.0&1.0&3.0&3.0\\
		$\eta^{\prime}$ mass window &...&...&0.7&0.7\\
		2D fit                             &3.9&1.2&4.8&4.8\\
		\hline
		$\gamma$ selection                 &...&...&...&1.0\\
		$\pi^0\,(\eta)$ reconstruction     &2.0&2.0&2.0&...\\
		Quoted $\mathcal B$                &0.03&0.4&1.3&1.4\\
		$\Delta E^{\rm sig}$ cut           &1.1&0.4&0.1&0.1\\
		MC statistics                      &0.2&0.2&0.3&0.2\\
		$K^0_S$ veto                       &...&...&...&1.8\\
		Fit bias                       &3.0&...&...&...\\
		Opening angle requirement                       &0.3&0.2&1.3&0.3\\
		
		Total                              &5.5&2.5&6.0&5.9\\
		\hline
	\end{tabular}
}
\end{table*}

\section{Summary}
\begin{figure*}
	\centering
	\subfigure[]{\includegraphics[width=0.55\linewidth]{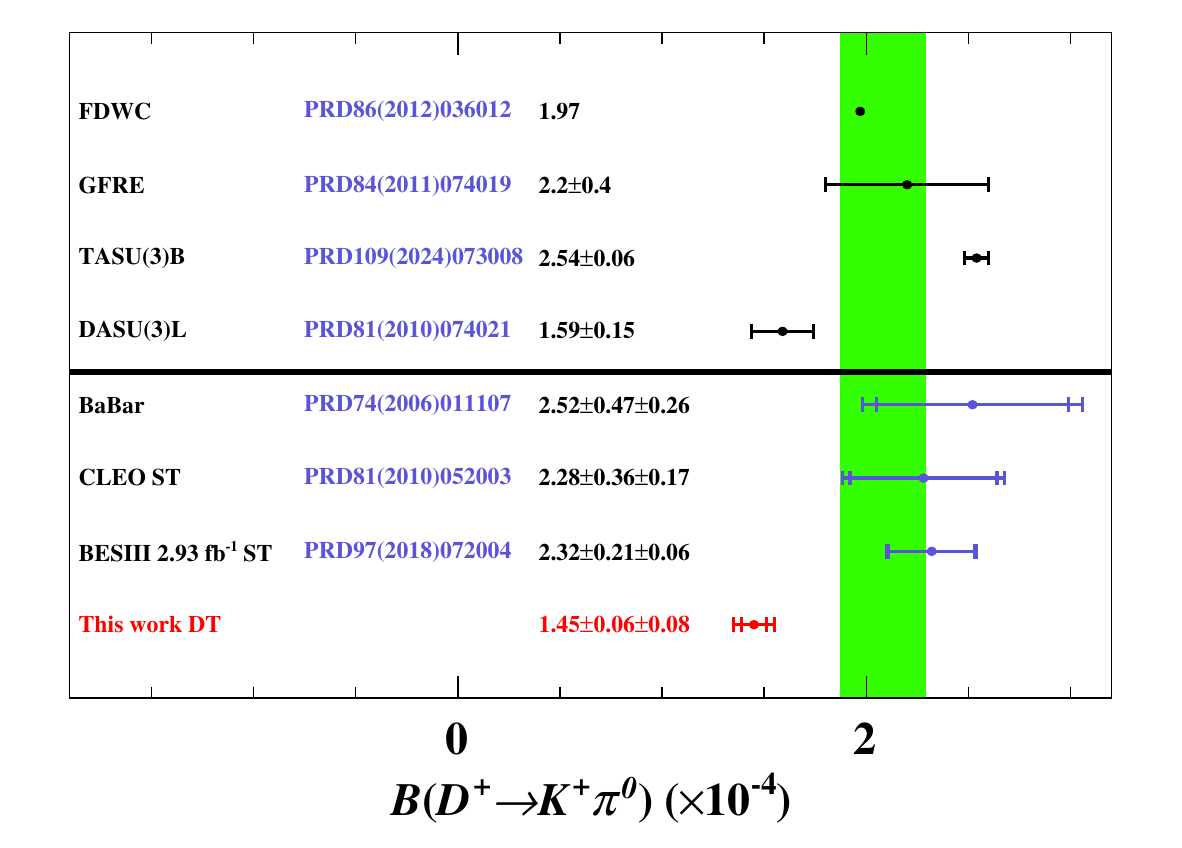}}
	\subfigure[]{\includegraphics[width=0.55\linewidth]{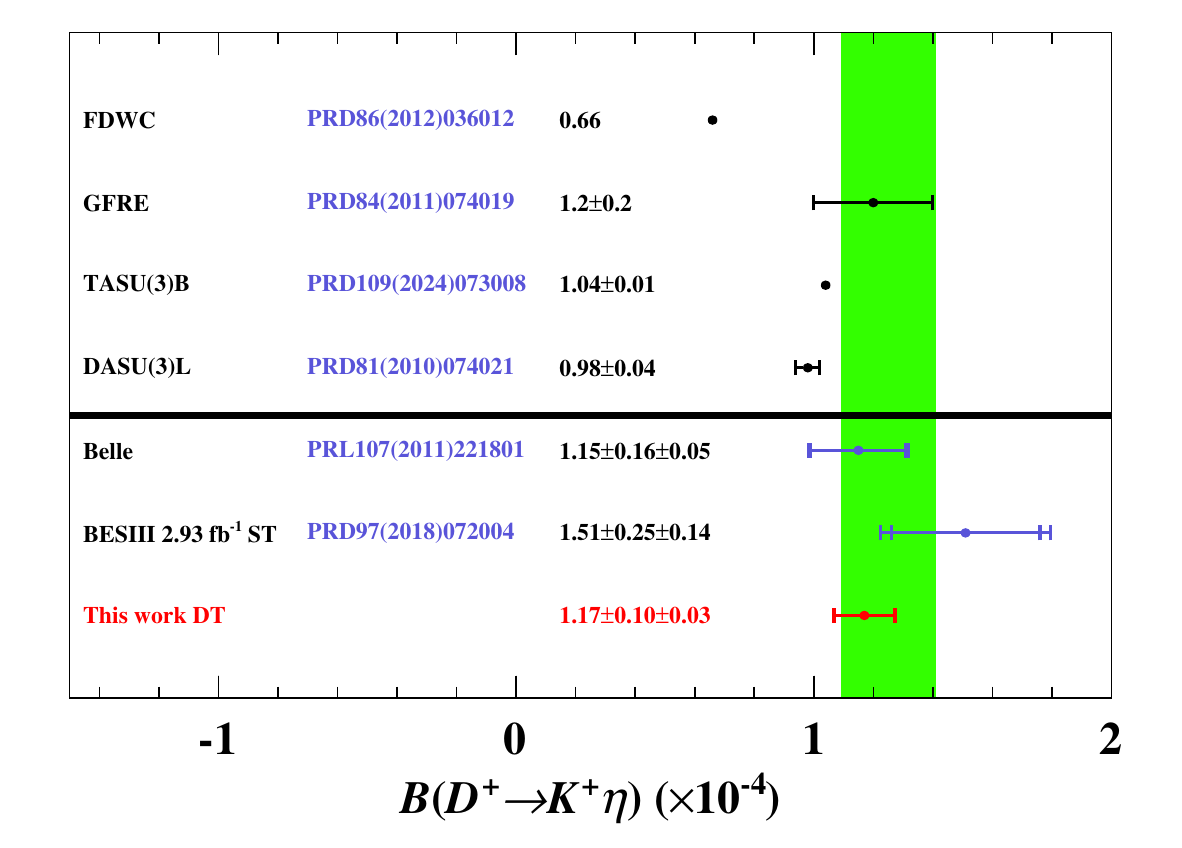}}
	\subfigure[]{\includegraphics[width=0.55\linewidth]{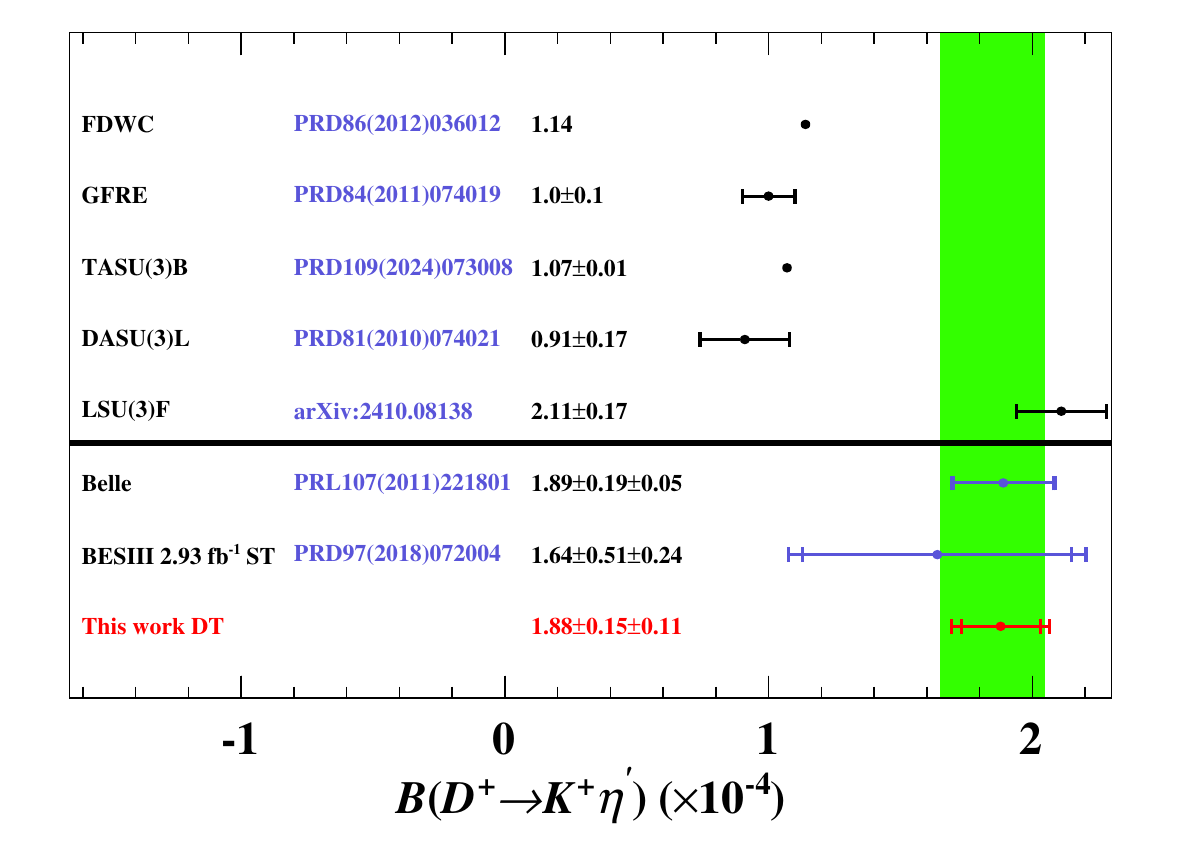}}
	\caption{
			Comparisons of experimental measurements and theoretical calculations for (a) $D^+\to K^+\pi^0$, (b) $D^+\to K^+\eta$ and (c) $D^+\to K^+\eta^{\prime}$. The green bands represent the $1\sigma$ PDG values. Results before and after the horizontal black line are theoretical and experimental results, respectively. For experimental results, the first uncertainties are statistical and the second ones are systematic. For theoretical calculations, the uncertainties are systematic only. }
	\label{fig:com}
\end{figure*}

In summary, using $20.3\,\rm fb^{-1}$ of $e^+e^-$ collision data taken at $\sqrt{s}=3.773$\,GeV with the BESIII detector, we measure the BFs of the DCS decays $D^+\to K^+\pi^0$, $D^+\to K^+\eta$ and $D^+\to K^+\eta^{\prime}$ to be ${\mathcal B}(D^+\to K^+ \pi^0) = (1.45 \pm 0.06 \pm 0.08)\times 10^{-4}$, ${\mathcal B}(D^+\to K^+ \eta) = (1.17 \pm 0.10 \pm 0.03)\times 10^{-4}$ and ${\mathcal B}(D^+\to K^+ \eta^{\prime}) = (1.88 \pm 0.15 \pm 0.11)\times 10^{-4}$, where the first uncertainties are statistical and the second are systematic. The precision of the BF of $D^+\to K^+ \pi^0$ is improved by a factor of two compared to the world average, while the latter two are comparable with the world averages. We notice that there are about four standard deviations between ${\mathcal B}(D^+\to K^+ \pi^0)$ measured in this work and that in Ref. \cite{ref:bes3all}. This difference is likely attributable to that the measurement in Ref. \cite{ref:bes3all} used the single-tag method, which suffers from much higher background, and the systematic uncertainty due to background shape could be substantially underestimated.~The BF of $D^+\to K^+\pi^0$ is consistent with the DASU(3)L and GFRE calculations within 3$\sigma$~\cite{ref:the1,ref:the11}, but disfavors with the TASU(3)B and FDWC calculations~\cite{ref:the4,ref:the12}. The BF of $D^+\to K^+\eta$ is consistent with the DASU(3)L, TASU(3)B and GFRE calculations within 3$\sigma$~\cite{ref:the1,ref:the4,ref:the11}, but disfavors with the FDWC  calculation~\cite{ref:the12}. The BF of $D^+\to K^+\eta^{\prime}$ is consistent with the LSU(3)F calculation within 3$\sigma$~\cite{ref:newinput}, but disfavors with the DASU(3)L, TASU(3)B, GFRE, and FDWC  calculations~\cite{ref:the1,ref:the4,ref:the11,ref:the12}. These new BFs are helpful to improve theoretical calculations. The comparisons of our results with other experimental measurements and theoretical calculations are shown in Table~\ref{tab:theory} and Fig.~\ref{fig:com}.

\begin{table*}[htp]
	\centering
	\caption{
		The BFs$~(\times 10^{-4})$ of $D^+\to K^+\pi^0$, $ D^+ \to K^+ \eta$ and $D^+\to K^+\eta^\prime$ from experimental measurements and theoretical calculations. For the experimental results, the first uncertainties are statistical and the second ones are systematic. For the theoretical predictions, the uncertainties are systematic only. The ...  means not measured}
	\label{tab:theory}
	\centering
%	\resizebox{1.0\textwidth}{!}{
	\begin{tabular}{cccc}
		\hline		
		Signal decay & $D^{+}\to K^{+}\pi^{0}$  & $D^{+}\to K^{+}\eta$ & $D^{+}\to K^{+}\eta^{\prime}$ \\
		\hline
		CLEO   \cite{ref:cleo}& $2.28\pm0.36\pm0.17$ & ... & ... \\
		Belle  \cite{ref:belle}& ... & $1.15\pm0.16\pm0.05$& $1.87\pm0.19\pm0.05$\\
		BaBar  \cite{ref:babar}& $2.52\pm0.47\pm0.26$& ... & ... \\
		BESIII \cite{ref:bes3all}& $2.32\pm0.21\pm0.06$& $1.51\pm0.25\pm0.14$& $1.64\pm0.51\pm0.24$\\
		PDG    \cite{ref:pdg2022}& $2.08\pm0.21$& $1.25\pm0.16$& $1.85\pm0.20$\\            
		DASU(3)L	\cite{ref:the1}   & $1.59\pm 0.15$ &$0.98\pm 0.04$ &$0.91\pm 0.17$ \\
		TASU(3)B \cite{ref:the4}  & $2.54\pm 0.06$& $1.04\pm 0.01$&$1.07\pm 0.01$\\
		GFRE    \cite{ref:the11} & $2.2\pm 0.4$& $1.2\pm 0.2$& $1.0\pm 0.1$\\
		FDWC \cite{ref:the12}   & 1.97& 0.66& 1.14\\  
		LSU(3)F	\cite{ref:newinput}   & ... &... &$2.11\pm 0.17$ \\         
		This work & $1.45\pm0.06\pm0.08$& $1.17 \pm 0.10 \pm 0.03$& $1.88 \pm 0.15\pm 0.11$\\ 
		\hline		
	\end{tabular}
%}
\end{table*}

\section{Acknowledgement}

The BESIII Collaboration thanks the staff of BEPCII (https://cstr.cn/31109.02.BEPC) and the IHEP computing center for their strong support. This work is supported in part by National Key R\&D Program of China under Contracts Nos. 2023YFA1606000, 2023YFA1606704; National Natural Science Foundation of China (NSFC) under Contracts Nos. 12035009, 11635010, 11935015, 11935016, 11935018, 12025502, 12035013, 12061131003, 12192260, 12192261, 12192262, 12192263, 12192264, 12192265, 12221005, 12225509, 12235017, 12361141819; The Chinese Academy of Sciences (CAS) Large-Scale Scientific Facility Program; CAS under Contract No. YSBR-101; 100 Talents Program of CAS; The Institute of Nuclear and Particle Physics (INPAC) and Shanghai Key Laboratory for Particle Physics and Cosmology; Agencia Nacional de Investigacin y Desarrollo de Chile (ANID), Chile under Contract No. ANID PIA/APOYO AFB230003; German Research Foundation DFG under Contract No. FOR5327; Istituto Nazionale di Fisica Nucleare, Italy; Knut and Alice Wallenberg Foundation under Contracts Nos. 2021.0174, 2021.0299; Ministry of Development of Turkey under Contract No. DPT2006K-120470; National Research Foundation of Korea under Contract No. NRF-2022R1A2C1092335; National Science and Technology fund of Mongolia; Polish National Science Centre under Contract No. 2024/53/B/ST2/00975; Swedish Research Council under Contract No. 2019.04595; U. S. Department of Energy under Contract No. DE-FG02-05ER41374.

%\clearpage
%\include{authorlist_2025-03-14}

\end{document}